# A deep *ROSAT* survey - IV. The evolution of X-ray-selected QSOs

B.J.Boyle,[1] T.Shanks,[2] I.Georgantopoulos,[3] G.C.Stewart[3] and R.E.Griffiths[4]

[1] *Institute of Astronomy, University of Cambridge, Madingley Road, Cambridge, CB3 0HA*
[2] *Department of Physics, University of Durham, South Road, Durham, DH1 3LE*
[3] *Department of Physics and Astronomy, The University, University Road, Leicester, LE1 7RH*
[4] *Department of Physics and Astronomy, Johns Hopkins University, Homewood Campus, Baltimore, MD 21218, USA*

**ABSTRACT**
We report on a new estimate of the QSO X-ray luminosity function and its evolution with redshift based on a sample of 107 QSOs detected at faint X-ray fluxes, $S(0.5 - 2 \,\text{keV}) > 4 \times 10^{-15} \,\text{erg s}^{-1}\text{cm}^{-2}$, with the *ROSAT* X-ray satellite. For $q_0 = 0.5$, the X-ray evolution of QSOs in this sample is consistent with strong luminosity evolution, $L^*_X(z) \propto (1+z)^{3.25\pm0.1}$, at low redshifts ($z < 1.60$) and a constant comoving space density at higher redshifts. The derived rate of evolution at low redshifts is thus significantly higher than that obtained previously for the *Einstein* Extended Medium Sensitivity Survey (EMSS). Indeed, most luminosity evolution models provide a very poor fit (rejected at the 99 per cent confidence level) when applied to the combined EMSS and *ROSAT* samples, although a polynomial evolution model, $L^*_X(z) \propto 10^{(1.14z - 0.23 z^2)}$, provides an adequate fit for $q_0 = 0$. For $q_0 = 0.5$, a simple power-law luminosity evolution model with a redshift cut-off ($L^*_X(z) \propto (1+z)^{2.51\pm0.1}$, $z_{\max} = 1.25$) is an acceptable fit to the EMSS and *ROSAT* samples only if a sizeable dispersion in the QSO X-ray spectral index ($\sigma(\alpha_X) \sim 0.5$) is included. More complex evolutionary forms, such as a variable luminosity function slope or density evolution combined with luminosity evolution, also fail to provide an adequate fit to the combined data-set. Possible systematic differences between the *Einstein* and *ROSAT* energy bands (e.g. spectral index variations) may account for some of the observed discrepancy between the QSO samples. Based on the observed range in the parameter values for the best-fit evolutionary models, we obtain formal values of 34–53 per cent for the QSO contribution to the 1–2 keV X-ray background.

## 1 INTRODUCTION

The QSO X-ray luminosity function (XLF) and its evolution with redshift provide fundamental information on the X-ray properties of QSOs. There are now a significant number of deep X-ray pointings which have been obtained with *ROSAT* (Shanks et al. 1991; Hasinger et al. 1993; Branduardi-Raymont et al. 1993), and in this paper we present the results of analysis of the QSO XLF and its evolution based on a sample of 107 QSOs detected to an X-ray flux limit $S(0.5 - 2 \,\text{keV}) > 4 \times 10^{-15} \,\text{erg s}^{-1}\text{cm}^{-2}$ with the Position Sensitive Proportional Counter (PSPC) on board *ROSAT* (see Georgantopoulos et al. 1994). In an earlier paper in this series (Boyle et al. 1993, hereafter Paper I) we reported on an initial determination of the QSO XLF, $\Phi_X(L_X, z)$, based on a preliminary sample of 42 QSOs detected on two deep ($\sim 27\,000$ s) *ROSAT* fields with an X-ray flux limit of $S(0.5 - 2\,\text{keV}) \gtrsim 8 \times 10^{-15} \,\text{erg s}^{-1}\text{cm}^{-2}$. We confirmed that the XLF exhibits a two-power-law form with a 'break' luminosity at $L_X(0.3 - 3.5\,\text{keV}) = 10^{43.9\pm0.1} \,\text{erg s}^{-1}$, first identified by Maccacaro et al. (1991, hereafter M91) and Della Ceca et al. (1992) from an analysis of 448 QSOs in the *Einstein* Extended Medium Sensitivity Survey (EMSS). We also confirmed that the evolution of the XLF was well represented by a power-law increase in luminosity with redshift $L_X \propto (1+z)^k$, but derived a higher rate for the combined EMSS/*ROSAT* samples ($k = 2.8 \pm 0.1$) than had been obtained by M91 ($k = 2.56$) based solely on an analysis of the EMSS. In addition, we found tentative evidence for a 'cut-off' in the luminosity evolution of QSOs at $z \gtrsim 2$, but had insufficient QSOs at these redshifts to establish firmly its existence or nature. Uncertainties in the evolution of QSOs at high redshift, coupled with the uncertainty in the value of the faint-end slope of the XLF, led to rather weak constraints (30–90 per cent) on the overall contribution of QSOs to the soft (2 keV) X-ray background.

In an attempt to improve our knowledge of the XLF and its evolution (particularly at low luminosities and at high redshifts) we have recently increased the size of our sample to 107 QSOs (Georgantopoulos et al. 1994) by obtaining fibre-optic spectroscopic identification for the optical counterparts to faint X-ray sources detected in a further three *ROSAT* fields, the deepest of which extends our X-ray flux limit to $S(0.5 - 2\,{\rm keV}) > 4 \times 10^{-15}\,{\rm erg\,s^{-1}\,cm^{-2}}$. In this paper we present the results of a new analysis of the XLF based on this larger sample of X-ray-selected QSOs. In Section 2 we present brief details of the QSO sample used in this paper, and we describe our analysis and results in Section 3. We discuss these results in Section 4 and present our conclusions in Section 5.

## 2   DATA

### 2.1   *ROSAT* observations

We base our analysis of the QSO XLF on a new sample of 107 QSOs detected with the PSPC on board *ROSAT* and identified spectroscopically using the AUTOFIB fibre-optic system at the Anglo-Australian Telescope (AAT). Full details of the X-ray and optical observations are presented elsewhere (Georgantopoulos et al. 1994; Shanks et al. in preparation), and thus only brief details will be given here. The X-ray sample is based on 5 deep PSPC exposures of fields originally surveyed spectroscopically for QSOs as part of the Durham-UVX QSO survey (Boyle et al. 1990). The names, centres and exposure times for each PSPC field are given in Table 1. The position of the deepest field, GSGP4, is offset by 10 arcmin from the position of the SGP4 field centre in the Durham-UVX QSO sample. This was done to allow the GSGP4 field to overlap with a deep galaxy redshift survey (Broadhurst, Ellis & Shanks 1988) also made in this region.

Over all 5 fields, 194 X-ray sources were detected at the $5\sigma$ limit in the 0.5–2 keV band. We use this band (rather than the full 0.1–2 keV band of the PSPC) to minimize any contribution from Galactic emission which can dominate below 0.5 keV. Due to the rapid increase in the size of the point-spread function (PSF) with off-axis angle, we limited our X-ray source detection to within 18 arcmin of the field centre in each PSPC image. This ensured that the PSF full width at half maximum intensity (FWHM) for any image is always less than two 15-arcsec pixels, and also avoided any significant obscuration by the telescope rib support structure which occurs at larger off-axis angles in the PSPC images. The total area surveyed at the $5\sigma$ limit is therefore equal to 1.41 deg$^2$. The flux limits corresponding to this detection limit not only vary from field-to-field (due to the different exposure times) but also vary within each field (due to the increase in the FWHM of the PSF with off-axis angle). To calculate the area covered by this survey as a function of X-ray flux, we therefore divided each field into 5 concentric annuli ($\theta \leq 10$ arcmin, $10 < \theta \leq 12$ arcmin, $12 < \theta \leq 14$ arcmin, $14 < \theta \leq 16$ arcmin and $16 < \theta \leq 18$ arcmin) and determined the flux limit in each annulus based on the field exposure time and mean theoretical FWHM of the PSF at this point. The resulting area coverage of the survey as a function of X-ray flux limit is given by the first two columns of Table 2.

**Table 1.** *ROSAT* PSPC fields.

| Field | RA (J2000) Dec | Exposure Time (secs) |
|---|---|---|
| SGP2  | 00 52 04.8   −29 05 24 | 24494 |
| SGP3  | 00 55 00.0   −28 19 48 | 21062 |
| GSGP4 | 00 57 28.7   −27 38 24 | 48955 |
| QSF1  | 03 42 09.6   −44 54 36 | 26155 |
| QSF3  | 03 42 14.3   −44 07 48 | 27358 |

### 2.2   Optical observations

COSMOS and APM measurements of the UK Schmidt $J$ plates in the South Galactic Pole region (J9771) and Field 249 (J2762) were used to identify the optical counterparts to the X-ray sources in the survey. A transformation between the *ROSAT* X-ray and COSMOS/APM optical co-ordinate frames was achieved using the positions of the existing 10–12 Durham-UVX survey QSOs detected by *ROSAT* on each field (see Shanks et al. 1991). Optical counterparts to each X-ray source were then identified out to a radius of 45 arcsec from the transformed X-ray position. At the plate limit ($B < 22.5$ mag), the majority of X-ray sources had at least one optical counterpart within 20 arcsec.

Low-resolution (12 Å) optical spectra were obtained for the nearest optical counterpart to each X-ray source using the AUTOFIB multi-object fibre-optic system at the AAT. In addition, we also obtained a few spectra for some of the fainter optical counterparts using the Low Dispersion Survey Spectrograph (LDSS) also at the AAT. In 5 nights we obtained optical identifications for the counterparts to 145 of the 194 X-ray sources. Of the 49 sources with no optical identification, in 30 cases the spectra were too poor to permit a reliable identification. For the remaining 19 objects, restrictions on the positioning of the fibres prevented the observation of the optical counterpart. The spectroscopic incompleteness was not independent of X-ray flux, ranging from $< 10$ per cent at $S(0.5 - 2\,{\rm keV}) > 4 \times 10^{-14}\,{\rm erg\,s^{-1}\,cm^{-2}}$ to $\sim 30$ per cent at $S(0.5 - 2\,{\rm keV}) > 4 \times 10^{-15}\,{\rm erg\,s^{-1}\,cm^{-2}}$. In the analysis below, we have corrected for this incompleteness by using an 'effective' survey area at each flux limit. For each flux limit in the survey, this 'effective area' was simply obtained by multiplying the total survey area by the fraction of sources successfully identified to that flux limit. These effective survey areas are also listed in Table 2.

Of the 145 sources with a reliable spectroscopic identification, 107 (74 per cent) were classified as QSOs. These 107 QSOs thus comprise the *ROSAT* QSO sample which will form the basis for the analysis presented below. An object was classified as a QSO if one or more broad emission lines (FWHM $> 1000$ km s$^{-1}$) were present in the optical spectrum. Based on the observed emission lines, we were able to determine an unambiguous redshift for 85 per cent of the QSOs. A further 12 galaxies exhibiting narrow emission lines (FWHM $<$ 1000 km s$^{-1}$) were also identified in this survey. Without spectral coverage of the redshifted H$\alpha$/[NII] region (and in some cases also the redshifted H$\beta$/[OIII] region) it was difficult to determine the true nature of these galaxies (i.e. starbursts, Seyfert 2), although their strong stellar continua and weak emission lines are very similar to the properties of 'normal' late-type galaxies commonly seen in field galaxies redshift surveys at conducted at similar optical magnitude limits ($B \sim 21$ mag, see Broadhurst et al. 1988). A more detailed discussion of their properties is presented in Georgantopoulos et al. (1994), although the effect of including these galaxies in the *ROSAT* QSO sample is considered explicitly in section 3.2.2.

Table 2. Cumulative area coverage of *ROSAT* survey.

| Flux Limit $S(0.5-2\,\text{keV})\,\text{erg}\,\text{s}^{-1}\,\text{cm}^{-2}$ | Total area (deg$^2$) | Effective area (deg$^2$) |
|---|---|---|
| $0.32\times10^{-14}$ | 0.09 | 0.06 |
| $0.33\times10^{-14}$ | 0.13 | 0.09 |
| $0.49\times10^{-14}$ | 0.17 | 0.12 |
| $0.51\times10^{-14}$ | 0.26 | 0.18 |
| $0.52\times10^{-14}$ | 0.30 | 0.21 |
| $0.54\times10^{-14}$ | 0.38 | 0.27 |
| $0.56\times10^{-14}$ | 0.42 | 0.30 |
| $0.60\times10^{-14}$ | 0.48 | 0.35 |
| $0.63\times10^{-14}$ | 0.56 | 0.43 |
| $0.64\times10^{-14}$ | 0.60 | 0.45 |
| $0.66\times10^{-14}$ | 0.69 | 0.52 |
| $0.68\times10^{-14}$ | 0.73 | 0.56 |
| $0.75\times10^{-14}$ | 0.77 | 0.61 |
| $0.79\times10^{-14}$ | 0.82 | 0.64 |
| $0.92\times10^{-14}$ | 1.02 | 0.82 |
| $0.10\times10^{-13}$ | 1.07 | 0.84 |
| $0.11\times10^{-13}$ | 1.12 | 0.88 |
| $0.12\times10^{-13}$ | 1.18 | 0.95 |
| $0.14\times10^{-13}$ | 1.24 | 1.04 |
| $0.15\times10^{-13}$ | 1.30 | 1.09 |
| $0.17\times10^{-13}$ | 1.36 | 1.17 |
| $0.18\times10^{-13}$ | 1.41 | 1.22 |
| $0.30\times10^{-13}$ | 1.41 | 1.24 |
| $0.40\times10^{-13}$ | 1.41 | 1.31 |
| $0.65\times10^{-13}$ | 1.41 | 1.41 |

As in Paper I, we convert the 0.5–2 keV QSO fluxes to the *Einstein* 0.3-3.5 keV band using the relation:

$$S(0.3-3.5\,\text{keV}) = 1.8\,S(0.5-2\,\text{keV})$$

which is accurate to $\pm 2$ per cent for all X-ray spectral indices $0.6 < \alpha_\text{X} < 1.5$, $f_\nu \propto \nu^{-\alpha_\text{X}}$. This enables us to combine our sample of 107 QSOs with the sample of 448 QSOs identified in the *Einstein* EMSS by Stocke et al. (1991). The EMSS QSO sample includes the 21 additional 'expected' QSOs in the EMSS, introduced to correct for the effects of incompleteness. In contrast to Paper I, we choose to incorporate these QSOs in the analysis below in order to maintain consistency with the *ROSAT* QSO sample, which has also been corrected for incompleteness (albeit in a different fashion). We also note that the EMSS QSO sample also includes 31 ambiguous and uncertain QSOs (tables 8 and 10 in Stocke et al. 1991). These objects exhibit predominantly narrow emission lines in their optical spectra, but were included in the QSO sample either because of suspected broad emission in the Balmer lines or because of high ionization implied from the [OIII]/[OII] line ratios. Some of these objects may thus be similar to the narrow emission line galaxies which we have also identified in our survey but chosen to exclude from our QSO sample. In the analysis below we will therefore consider the EMSS QSO sample both with and without the inclusion of these ambiguous QSOs in order to ascertain the importance of these objects to the determination of the XLF and its evolution.

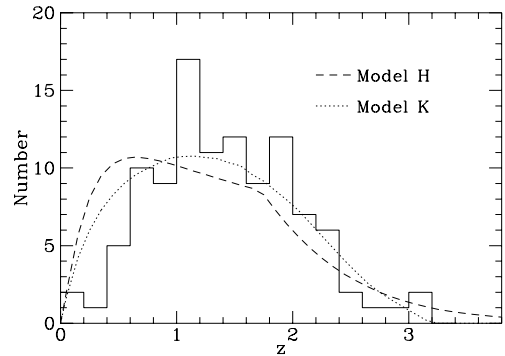

**Figure 1.** The number-redshift, $n(z)$, diagram for the 107 QSOs identified in the *ROSAT* sample. The dashed and dotted lines represent the $n(z)$ predictions for models H and K respectively (see text).

In Fig. 1 we present the number-redshift, $n(z)$, relation for the *ROSAT* sample. The $n(z)$ relation exhibits a broad maximum over the redshift range $0.5 < z < 1.8$. The median redshift for the sample is $z = 1.5$, greater than the median redshift for the EMSS, $z = 0.2$. The X-ray luminosity-redshift $(L_\text{X}, z)$ diagram for the *ROSAT* and EMSS QSOs is shown in Fig. 2. The QSO X-ray luminosities have been calculated for $q_0 = 0.5$ and $H_0 = 50$ km s$^{-1}$Mpc$^{-1}$ in the rest frame 0.3–3.5 keV energy band, assuming a power-law spectrum, $f_\nu \propto \nu^{-\alpha_\text{X}}$, with spectral index $\alpha_\text{X} = 1$. This is the mean spectral index for radio-quiet QSOs in the EMSS (Wilkes & Elvis 1987). Although the *ROSAT* QSOs appear to exhibit a slightly softer mean spectrum, $\alpha_\text{X} = 1.2$ (Stewart et al., in preparation), in order to maintain consistency with previous analyses (Paper I; M91; Della Ceca et al. 1992), we retain the assumption that $\alpha_\text{X} = 1$ throughout this paper. For the power-law luminosity evolution models considered below ($L_\text{X} \propto (1+z)^k$, see Section 3.2.1), an increase in adopted value of $\alpha_\text{X}$ simply increases the derived rate of evolution, $k$, by the same amount.

In Paper I we also considered a two-component power-law model for the X-ray spectra of QSOs, which introduced significant softening (an increase in the value of $\alpha_\text{X}$) of the spectrum below 1.5 keV. With additional data and improved spectral fitting it is now apparent that, although the X-ray spectra of QSOs do indeed soften at low energies, this softening occurs below 0.5 keV in the rest-frame of the QSOs (Stewart et al., in preparation), and is thus unimportant for the calculation of rest-frame luminosities in this analysis. From Fig. 2, we see that the addition of the *ROSAT* sample extends coverage of the QSO $(L_\text{X}, z)$ plane by more than an order of magnitude in luminosity. Although the *ROSAT* sample provides little improvement in the statistics of low-redshift ($z < 0.2$), low-luminosity ($L_\text{X} < 10^{43}$erg s$^{-1}$)

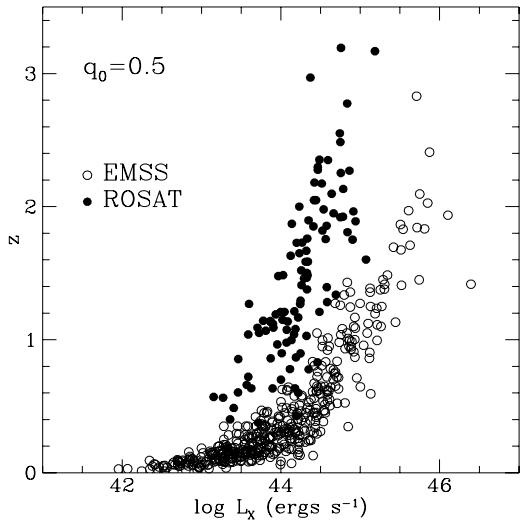

**Figure 2.** The X-ray luminosity-redshift diagram for the QSOs identified in the *ROSAT* sample (filled circles) and EMSS (open circles).

QSOs, it contains 80 QSOs with $z > 1$, almost twice as many QSOs as the EMSS (46 QSOs) over the same redshift range.

## 3 ANALYSIS

### 3.1  $1/V_a$ analysis

In order to obtain an initial estimate of the overall behaviour of the XLF and its dependence on redshift, we have derived binned estimates for the QSO XLF using the $1/V_a$ statistic (Avni & Bahcall 1980). This method is fully described by M91 and Paper I. In Figs 3(a) and 3(b) we plot the derived 0.3–3.5 keV QSO XLF for the combined *ROSAT* and EMSS (including ambiguous QSOs) samples for $q_0 = 0$ (Fig. 3a) and $q_0 = 0.5$ (fig 3b). The corresponding QSO XLFs based on the *ROSAT* sample alone are shown in Figs 3(c) and 3(d). In these figures, the XLFs are plotted in four redshift intervals of equal width in $\log(1 + z)$ over the range $0 < z < 3$. Error bars on these estimates of the XLF are obtained from Poissonian statistics. In this paper, we choose to quote the binned XLF per logarithmic luminosity interval rather than per linear luminosity interval adopted in Paper I and M91. The advantage of the approach adopted here is that luminosity evolution, in which the total number of QSOs is conserved with cosmic epoch, is represented by a purely horizontal shift of the XLF with redshift in this diagram. To avoid any misleading impression of the XLF coverage provided by these samples, all bins that contain three QSOs or less have been excluded from these figures.

From Figs 3(a) and 3(b), we see that the XLF exhibits a characteristic two-power-law form, first identified by M91 for the low-redshift ($z < 0.2$) XLF. In this case, however, the inclusion of the *ROSAT* sample has extended the XLF to lower luminosities at $z > 0.2$ and revealed that this two-power-law form is also apparent at these redshifts. This confirms the initial result reported in Paper I. The contribution of the *ROSAT* sample to the XLF is also borne out by Figs 3(c) and 3(d), where it can be seen that most of the QSOs identified in the *ROSAT* sample lie around the 'break' luminosity or fainter for $0.4 < z < 2.9$. From these figures it is also apparent that the redshift dependence of the XLF is governed predominantly by strong luminosity evolution for $z \lesssim 1.8$. This strong evolution is confirmed by the $<V_e/V_a>$ statistic (Avni & Bahcall 1980) calculated both for the combined data-set ($<V_e/V_a> = 0.67 \pm 0.01$) and for the *ROSAT* sample alone ($<V_e/V_a> = 0.63 \pm 0.03$). Both values are significantly different from the no-evolution value, $<V_e/V_a> = 0.5$, at greater than the $4\sigma$ level. The approximately constant luminosity shift between successive $\Delta \log(1 + z)$ redshift bins at $z < 1.8$ suggests that the evolution could take a power-law form, $L_X \propto (1 + z)^k$, at these redshifts. However, care should be taken not to over-interpret any binned representation of the XLF, as a significant amount of evolution can occur within each redshift interval plotted in such diagrams. Moreover, for luminosity evolution, even the higher value for $<V_e/V_a>$ (which primarily tests for density evolution) obtained for the combined sample compared to that derived for the *ROSAT* sample should not be taken to imply that the *ROSAT* sample exhibits less evolution than the EMSS. Indeed, the lower value simply reflects the fact that the *ROSAT* sample lies mostly on the flatter part of the XLF.

### 3.2  Maximum likelihood analysis

#### 3.2.1  Models

In order to derive a more accurate parametric representation of the QSO XLF and its evolution, we turn to the technique of maximum likelihood analysis employed in Paper I. This method, developed by Marshall et al. (1984) for use with the QSO optical luminosity function, yields best-fit values for parameters in a functional model used to describe the XLF and its evolution. As in Paper I, we use the two-power-law form for the $z = 0$ XLF proposed by M91 (with modified normalization as described in Paper I):

$$\Phi_X(L_X) = \Phi_X^* L_{X_{44}}^{-\gamma_1} \qquad L_X < L_X^*(z=0)$$

$$\Phi_X(L_X) = \frac{\Phi_X^*}{L_{X_{44}}^{*(\gamma_1-\gamma_2)}} L_{X_{44}}^{-\gamma_2} \qquad L_X > L_X^*(z=0)$$

where $\Phi_X^*$ is the normalization of the XLF and $\gamma_1$, $\gamma_2$ are the faint- and bright-end slopes respectively of the XLF. $L_{X_{44}}$ is 0.3–3.5 keV X-ray luminosity expressed in units of $10^{44}$ erg s$^{-1}$. The evolution of the XLF is fully described by the redshift dependence of the 'break' luminosity, $L_X^*(z)$. As in Paper I, we investigate the two most commonly used functional forms to describe the evolution: power-law evolution in $(1 + z)$,

$$L_X^*(z) = L_X^*(0) (1 + z)^k,$$

and exponential evolution with fractional look-back time ($\tau$),

$$L_X^*(z) = L_X^*(0) \exp(k\tau).$$

We also consider power-law evolution models in which the evolution 'switches off' at some maximum redshift $z_{\max}$:

$$L_X^*(z) = L_X^*(z_{\max}) \qquad z > z_{\max}.$$

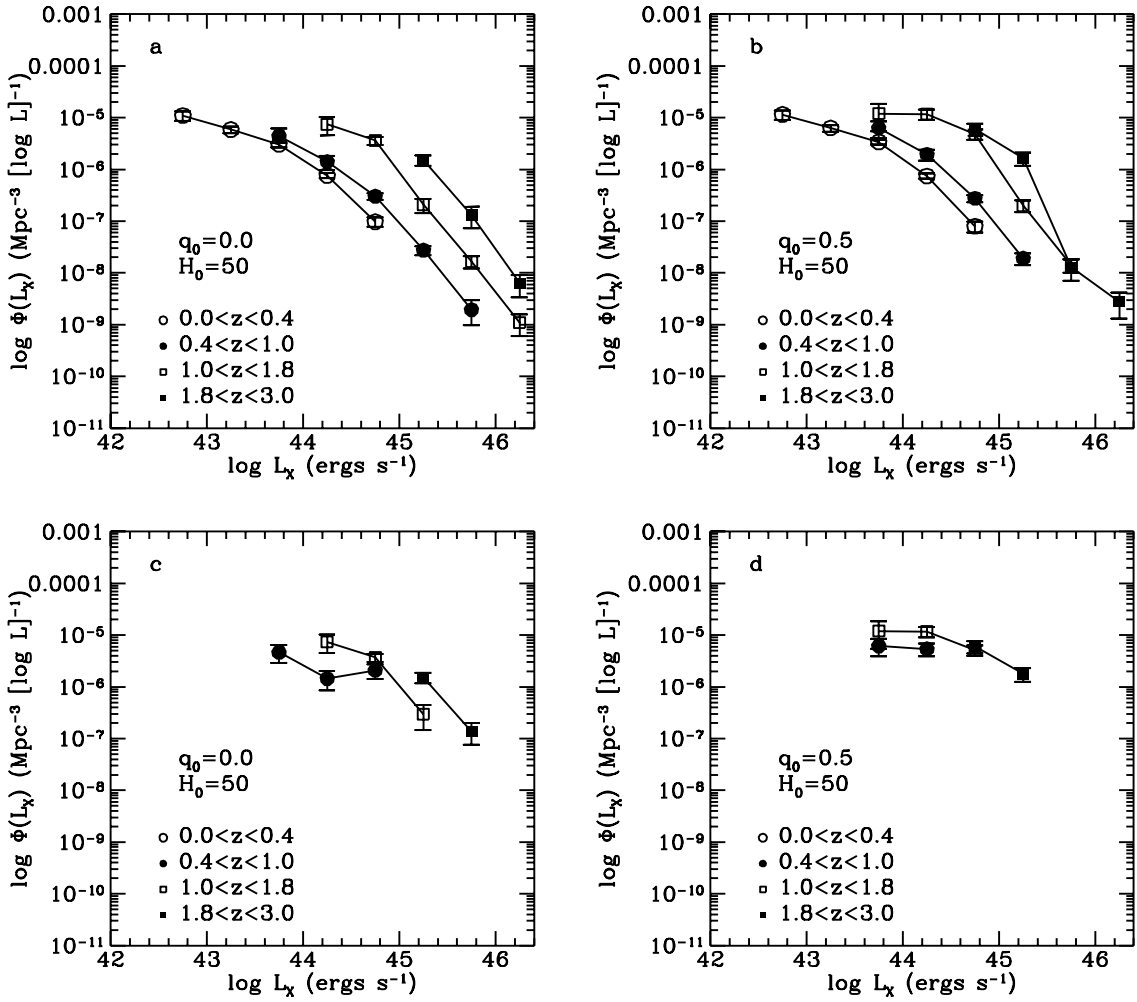

**Figure 3.** Binned $1/V_a$ estimates for the QSO X-ray luminosity function in the 0.3–3.5 keV band based on: (a) EMSS and *ROSAT* samples for $q_0 = 0.0$, (b) as (a) for $q_0 = 0.5$, (c) *ROSAT* sample only for $q_0 = 0.0$, d) as (c) for $q_0 = 0.5$.

Beyond $z_{\max}$ the XLF is independent of redshift and the comoving space density of QSOs remains constant for any given X-ray luminosity. To complete the possible parametrizations of pure luminosity evolution, we also investigate a 'polynomial' evolution model,

$$L_X^*(z) = L_X^*(0) \, 10^{(kz + k_2 z^2)},$$

similar to that adopted by Dunlop & Peacock (1990) in their study of the QSO radio luminosity function.

In addition to pure luminosity evolution models, we have also investigated more complex evolutionary forms. We have tried to fit a model which incorporates both luminosity and density evolution. In this case the redshift dependence of the XLF is not only modelled by a $(1+z)$ power-law luminosity evolution (as above) but also includes a $(1+z)$ power-law evolution in comoving number density:

$$\Phi_X^*(z) = \Phi_X^*(0) \, (1+z)^{k_2}.$$

We have also studied models in which the shape of the XLF is dependent on redshift, and have considered a power-law luminosity evolution model in which the faint-end slope also varies linearly with $z$:

$$\gamma_1(z) = \gamma_1(0) + k_2 z.$$

Any potentially large dispersion in the X-ray spectral indices for QSOs could also affect the results derived in our analysis below. Giallongo & Vagnetti (1992) have demonstrated that any dispersion in the adopted spectral index for QSOs can also significantly reduce the derived rate of evolution. To estimate the size of this effect in the X-ray samples considered below, we have included fits in which the dispersion in X-ray spectral index, $\sigma(\alpha_X)$, is assumed to be $\sigma(\alpha_X) = 0.25$ and $\sigma(\alpha_X) = 0.5$. In order to calculate the best-fit model in these cases, we have used the expression derived by Francis (1993) to calculate the mean observed X-ray spectral index, $\alpha_{X_{obs}}$, as a function of redshift introduced as a result of the dispersion in the spectral index:

$$\alpha_{X_{obs}} = \alpha_X + \sigma^2 (1 - \gamma_2) \ln(1+z).$$

For an XLF steeper than $\gamma_2 = 1$ (as is observed here), any dispersion in the spectral index will therefore cause an ef-fective 'hardening' in the mean observed spectral index with redshift, resulting in a corresponding decrease in the value of the evolution rate $k$. Although, this expression only strictly applies to a single-power-law LF, most QSOs in this analysis lie on the steep portion of the XLF, and so the value of $\gamma_2$ can used for the slope without any significant loss of accuracy. Note that dispersion in spectral index does not affect the shape of the XLF and so the value of $\gamma_2$ used in this expression can be obtained from the value derived in the absence of any dispersion in the spectral index, without requiring any iteration of the above equation to obtain consistent values of $\alpha_{X_{obs}}$ and $\gamma_2$ in the model fits.

In the model fits to the XLF there can thus be four, five or six free parameters: $\gamma_1$, $\gamma_2$, $L_X^*(0)$, $k$, plus $z_{max}$ and/or $k_2$, depending on the evolutionary model chosen. The best-fit values for these parameters may be obtained from the maximum likelihood analysis. Statistical errors on these best-fit values (corresponding to 68 per cent confidence regions) were derived using the method described by Boyle, Shanks & Peterson (1988). However, the range in the values (particularly those for the $z = 0$ XLF) from model to model may give a better indication of the full uncertainty in these parameters. The normalization of the XLF ($\Phi_X^*$) is not a free parameter in the fit and its value is derived from the normalization of the best-fit model to the total number of QSOs observed in the samples (see Paper I).

Although the maximum likelihood analysis gives best-fit values for the model parameters, it does not give a 'goodness of fit' for the best-fit model. To test the overall acceptability of the model fit to the data, we therefore used the two-dimensional (2D) KS statistic (Peacock 1985) employing the algorithm devised by Press et al. (1992). The KS probability ($P_{KS}$) is derived from comparison of the two-dimensional cumulative probability distributions in luminosity and redshift for both the model and data. As in Paper I, these distributions were tested over the full luminosity and redshift range of the combined sample, $10^{42} < L_X < 10^{48}$ erg s$^{-1}$, $0 < z < 3$.

### 3.2.2 Results

The results of the maximum likelihood analysis and 2D KS tests applied to the combined EMSS and *ROSAT* samples are presented in Table 3 for a variety (a veritable A-Z) of the models discussed above. With the inclusion of the larger *ROSAT* sample in this paper, we now find that neither of the simplest evolutionary forms (exponential or power-law without a redshift cut-off) provides an acceptable fit (at the 99 per cent confidence level) to the combined EMSS and *ROSAT* samples for either $q_0 = 0$ or $q_0 = 0.5$ (models A-D). However, the KS probabilities of both the exponential and power-law evolution models are higher for the $q_0 = 0$ case than for $q_0 = 0.5$, in agreement with the trend found in Paper I. Allowance for a dispersion in the X-ray spectral index for QSOs does not increase the acceptability of these simple models. Indeed, the inclusion of an increasingly large dispersion in the spectral index for the $q_0 = 0.5$ power-law evolution model (models E and F) decreases the overall acceptability of the fit.

Similarly, the inclusion of a cut-off in the evolution at high redshift does not result in an acceptable model fit to the data. For $q_0 = 0$, the addition of a redshift cut-off (model G) only results in a very small increase in the overall KS probability of the model fit (from 0.5 per cent to 0.7 per cent). In contrast, for $q_0 = 0.5$, the inclusion of a redshift cut-off in the power-law evolution model does significantly increase the acceptability of the model fit, from 0.02 per cent (model C) to 0.5 per cent (model H). Nevertheless, at this KS probability, this model can hardly be considered a good fit to the data. A better fit can be achieved if the X-ray spectral index is allowed to have a significant dispersion (models I and J). For a $\sigma(\alpha_X) = 0.5$, the KS probability for the $q_0 = 0.5$, power-law evolution ($k = 2.51$) with redshift cut-off ($z_{max} = 1.25$) is 1.3 per cent, formally acceptable at the 1 per cent level.

A similar level of acceptability, without the inclusion of a dispersion in the spectral index, is found for the polynomial evolution model in a $q_0 = 0$ universe (model K), although the corresponding model for $q_0 = 0.5$ (model L) has a much lower KS probability (0.3 per cent). Inclusion of additional coefficients in the polynomial model (i.e. a third-order fit) yields no improvement in the overall acceptability of the model for $q_0 = 0.5$. Indeed the formal best-fit value for the $z^3$ coefficient in a cubic polynomial fit to the data in a $q_0 = 0.5$ universe is $k_3 = 0$. It is intriguing that the values for the evolution parameters $k$ and $k_2$ in the polynomial evolution model, $k = 1.14(1.15)$ and $k_2 = -0.23(-0.26)$ for $q_0 = 0.0(0.5)$, are remarkably similar to those found by Dunlop & Peacock (1990), $k = 1.18$, $k_2 = -0.28$ for the radio evolution of flat-spectrum radio sources (predominantly QSOs) in a $q_0 = 0.5$ universe.

Given the poor fit of many of the luminosity evolution models applied to the data, we also considered the more complicated evolutionary models discussed above. However, as can be seen from Table 3, neither the luminosity and density evolution models (models M and N) nor the models incorporating a variable faint end slope (models O and P) provide improved fits to the combined data-set in either $q_0 = 0$ or $q_0 = 0.5$ universes.

Although we have accounted for X-ray flux-dependent incompleteness in the *ROSAT* survey (see Section 2.2), it is possible that some optical magnitude-dependent incompleteness has also been introduced by the decreasing success rate for optical identifications at faint magnitudes in the spectroscopic survey. Our correction for incompleteness (i.e. effective areas) assumes that these unidentified sources contain the same relative numbers of source types (QSOs, galaxies, etc.) as the identified sources. Since QSOs are generally the easiest class of object to identify in low signal-to-noise ratio spectra, this assumption may not be strictly correct. Currently 49 out of 194 objects remain unidentified, of which 30 are as a result of poor optical spectra, corresponding to spectroscopic incompleteness of 17 per cent. In order to establish whether any possible spectroscopic bias in this population significantly influences our results, we have therefore carried out the following two tests.

First, we increased the X-ray flux limit in the *ROSAT* sample to $S(0.3 - 3.5 \text{ keV}) > 2.0 \times 10^{-14}$ erg s$^{-1}$ cm$^{-2}$, corresponding to the point at which the spectroscopic incompleteness falls to below 10 per cent. At this limit, the *ROSAT* sample contains 105 X-ray sources (including 71 QSOs), with 9 objects unidentified due to poor optical spectra and a further 10 sources not observed due to fibre positioning restrictions. The results of performing the maximum likelihood analysis for the power-law evolution model with redshift cut-off on this restricted $ROSAT$ sample, combined with the EMSS, are listed in Table 3 (model H′). Compared with the corresponding model for the full $ROSAT$ sample (model H) we can see that the use of the restricted sample has made no significant difference (at $> 2\sigma$ level) to the overall model parameters, or to the acceptability of the fit.

Table 3. QSO XLF models.

| Model | $q_0$ | Evolution | $\sigma(\alpha_X)$ | $\gamma_1$ | $\gamma_2$ | $\log L_X^*(0)$† | $k$ | $z_{max}$ | $k_2$ | $\Phi_X^*$‡ | $P_{KS}$ |
|---|---|---|---|---|---|---|---|---|---|---|---|
| A | 0.0 | $(1+z)^k$ | 0.00 | 1.58 | 3.35 | 43.81 | 2.63 | | | 0.72 | $4.8 \times 10^{-3}$ |
| B | 0.0 | $exp(k\tau)$ | 0.00 | 1.39 | 3.30 | 43.66 | 5.10 | | | 0.74 | $3.5 \times 10^{-3}$ |
| C | 0.5 | $(1+z)^k$ | 0.00 | 1.12 | 3.31 | 43.29 | 2.44 | | | 1.35 | $1.6 \times 10^{-4}$ |
| D | 0.5 | $exp(k\tau)$ | 0.00 | 1.14 | 3.33 | 43.27 | 4.02 | | | 2.34 | $1.7 \times 10^{-3}$ |
| E | 0.5 | $(1+z)^k$ | 0.25 | 1.41 | 3.29 | 43.70 | 2.20 | | | 1.30 | $2.6 \times 10^{-5}$ |
| F | 0.5 | $(1+z)^k$ | 0.50 | 1.44 | 3.29 | 43.79 | 1.54 | | | 1.24 | $4.3 \times 10^{-7}$ |
| G | 0.0 | $(1+z)^k + z_{max}$ | 0.00 | 1.53 | 3.38 | 43.70 | 3.03 | 1.89 | | 0.79 | $6.7 \times 10^{-3}$ |
| H | 0.5 | $(1+z)^k + z_{max}$ | 0.00 | 1.36 | 3.37 | 43.57 | 2.90 | 1.73 | | 1.45 | $4.8 \times 10^{-3}$ |
| H′ | 0.5 | $(1+z)^k + z_{max}$ | 0.00 | 1.37 | 3.53 | 43.54 | 3.11 | 1.83 | | 1.64 | $2.7 \times 10^{-3}$ |
| H″ | 0.5 | $(1+z)^k + z_{max}$ | 0.00 | 1.36 | 3.32 | 43.59 | 2.80 | 1.75 | | 1.40 | $3.2 \times 10^{-3}$ |
| I | 0.5 | $(1+z)^k + z_{max}$ | 0.25 | 1.45 | 3.44 | 43.61 | 2.93 | 1.45 | | 1.12 | $5.4 \times 10^{-3}$ |
| J | 0.5 | $(1+z)^k + z_{max}$ | 0.50 | 1.46 | 3.45 | 43.65 | 2.51 | 1.25 | | 1.12 | $1.3 \times 10^{-2}$ |
| K | 0.0 | $kz + k_2 z^2$ | 0.00 | 1.50 | 3.35 | 43.71 | 1.14 | | −0.23 | 0.84 | $1.5 \times 10^{-2}$ |
| L | 0.5 | $kz + k_2 z^2$ | 0.00 | 1.26 | 3.32 | 43.52 | 1.15 | | −0.26 | 1.89 | $2.6 \times 10^{-3}$ |
| M | 0.0 | $(1+z)^k + (1+z)^{k_2}$ | 0.00 | 1.53 | 3.35 | 43.78 | 2.76 | | −0.26 | 0.82 | $4.5 \times 10^{-3}$ |
| N | 0.5 | $(1+z)^k + (1+z)^{k_2}$ | 0.00 | 1.55 | 3.34 | 43.81 | 1.91 | | 1.03 | 0.78 | $1.2 \times 10^{-3}$ |
| O | 0.0 | $(1+z)^k + z_{max} + k_2 z$ | 0.00 | 1.46 | 3.39 | 43.65 | 3.16 | 1.92 | −0.18 | 0.98 | $5.8 \times 10^{-3}$ |
| P | 0.5 | $(1+z)^k + z_{max} + k_2 z$ | 0.00 | 1.39 | 3.39 | 43.63 | 2.77 | 1.60 | 0.25 | 1.16 | $2.9 \times 10^{-3}$ |
| Q§ | 0.0 | $(1+z)^k$ | 0.00 | 1.58 | 3.35 | 43.81 | 2.66 | | | 0.74 | $2.2 \times 10^{-2}$ |
| R§ | 0.5 | $(1+z)^k$ | 0.00 | 1.12 | 3.31 | 43.29 | 3.17 | | | 2.59 | $1.4 \times 10^{-5}$ |
| S§ | 0.0 | $(1+z)^k + z_{max}$ | 0.00 | 1.53 | 3.38 | 43.70 | 3.34 | 1.79 | | 0.63 | $3.1 \times 10^{-1}$ |
| T§ | 0.5 | $(1+z)^k + z_{max}$ | 0.00 | 1.36 | 3.37 | 43.57 | 3.25 | 1.60 | | 1.59 | $1.2 \times 10^{-2}$ |
| U¶ | 0.5 | $(1+z)^k + z_{max}$ | 0.00 | 1.25 | 3.40 | 43.52 | 3.17 | 1.48 | | 1.61 | $1.9 \times 10^{-3}$ |
| V¶¶ | 0.5 | $(1+z)^k + z_{max}$ | 0.00 | 1.59 | 3.44 | 43.63 | 2.92 | 1.75 | | 0.96 | $1.2 \times 10^{-2}$ |
| W+ | 0.0 | $(1+z)^k + z_{max}$ | 0.00 | 1.53 | 3.18 | 43.81 | 2.14 | 1.25 | | 0.76 | $6.3 \times 10^{-2}$ |
| X+ | 0.5 | $(1+z)^k + z_{max}$ | 0.00 | 1.02 | 3.02 | 43.41 | 2.49 | 0.95 | | 3.39 | $1.0 \times 10^{-2}$ |
| Y* | 0.0 | $(1+z)^k + z_{max}$ | 0.00 | 1.60 | 3.51 | 43.88 | 2.93 | 1.80 | | 0.95 | $2.0 \times 10^{-2}$ |
| Z* | 0.5 | $(1+z)^k + z_{max}$ | 0.00 | 1.57 | 3.81 | 43.79 | 3.13 | 1.58 | | 1.27 | $5.9 \times 10^{-2}$ |
| Errors | | | | ±0.15 | ±0.1 | ±0.2 | ±0.1 | ±0.1 | ±0.1 | | |

† In units of $erg\,s^{-1}$.
‡ In units of $10^{-6} Mpc^{-3} (10^{44}\,erg\,s^{-1})^{-1}$.
′ $ROSAT$ sample limited to $S(0.3 - 3.5\,keV) > 2 \times 10^{-14}\,erg\,s^{-1} cm^{-2}$.
″ Excluding all unidentified sources in the $ROSAT$ sample.
§$ROSAT$ sample only. Parameters for $z = 0$ XLF held fixed in fit.
¶ Excluding EMSS 'ambiguous' QSOs.
¶¶ Including $ROSAT$ narrow emission line galaxies.
+ No correction applied to $ROSAT$ 0.5–2 keV fluxes.
* EMSS sample limited to $S(0.3 - 3.5\,keV) > 2.8 \times 10^{-13}\,erg\,s^{-1} cm^{-2}$.

Secondly, we also performed the maximum likelihood analysis for the same evolution model on the full $ROSAT$ sample, assuming that none of the unidentified sources was a QSO (model H″). Although this assumption is almost certainly unrealistic (since it also includes sources that were simply not observed), it does provide an upper limit in the more likely case that the majority of the unidentified sources are objects with weak emission line or absorption spectra (e.g. galaxies, clusters of galaxies, BL Lacs) and not QSOs. However, as can be seen from the values of best-fit parameters for model H″ listed in Table 3, even this pessimistic assumption makes no significant difference to the overall result. From these two tests we therefore conclude that any optical magnitude effects introduced by the spectroscopic incompleteness do not significantly influence our results.

## 4 DISCUSSION

### 4.1 *ROSAT* versus EMSS evolution

It is clear from the previous section that few of the models investigated in this paper provide an acceptable fit to the combined EMSS and $ROSAT$ samples. Although pure luminosity evolution models can be found that will formally fit the data at the 1 per cent level for both $q_0 = 0$ (polynomial evolution, model K) and $q_0 = 0.5$ (power-law evolution with large dispersion in X-ray spectral index, model J), neither of the models is acceptable at the 5 per cent level. Indeed, more complex evolutionary forms (e.g. luminosity and density evolution, models M and N) provide no better fits to the combined data-set.

Nevertheless, simple evolution models do indeed provide good fits ($P_{KS} > 10$ per cent) to both EMSS and $ROSAT$ data-sets when the maximum likelihood analysis is performed separately on each sample. In Paper I, we demonstrated that the EMSS QSOs were consistent with power-law evolution, $L_X^*(z) \propto (1+z)^{2.56}$ for $q_0 = 0.5$, with the *ROSAT* sample (which then contained only 42 QSOs) exhibiting a much higher rate of evolution, $L_X^*(z) \propto (1+z)^{2.9}$. The rate of evolution was derived for the *ROSAT* sample by adopting the best-fit values for the parameters of the $z = 0$ XLF obtained for the combined EMSS and *ROSAT* samples ($\gamma_1$, $\gamma_2$ and $L_X^*(0)$) and deriving only the value of $k$ from the maximum likelihood analysis. If we adopt a similar procedure for the larger *ROSAT* data-set, we again find that we derive much higher rates for the evolution of the XLF, with $L_X^*(z) \propto (1+z)^{2.66}$ and $L_X^*(z) \propto (1+z)^{3.17}$ for $q_0 = 0$ (model Q) and $q_0 = 0.5$ (model R) respectively.

However, neither model is a good fit to the *ROSAT* data: the $q_0 = 0$ model is rejected at the 95 per cent confidence level, while the $q_0 = 0.5$ model is rejected at the 99 per cent confidence level. A much better fit to the *ROSAT* sample is obtained for models with a cut-off in the luminosity evolution at $z_{\max}$ (models S and T). From the maximum likelihood analysis, we obtain best-fit parameter values of $z_{\max} = 1.6(1.7)$ and $k = 3.25(3.34)$ for $q_0 = 0.5(0.0)$. Both fits have a much higher KS probability ($> 10$ per cent) than any of the other models and confirm that pure luminosity evolution with a redshift cut-off is an acceptable fit to the *ROSAT* sample.

The value for the evolution parameter ($k = 3.34 \pm 0.1$) derived from the *ROSAT* sample alone ($q_0 = 0.0$) is significantly higher than that derived from the EMSS QSO sample ($k = 2.56$), or even from the combined EMSS/*ROSAT* sample ($k = 3.03 \pm 0.1$). The high rate of evolution implied by the *ROSAT* sample is also apparent from the $n(z)$ diagram plotted in Fig. 1, where the observed median redshift for the sample is higher than than predicted by the power-law evolution fit (model H, dashed line) to the combined data-set. However, a significantly better fit to the observed $n(z)$ is achieved with the $q_0 = 0$ polynomial evolution model (model K, dotted line).

In Fig. 1 we have normalized both the predicted number-redshift relations (models H and K) to the number of QSOs observed in the *ROSAT* sample. However, it is a general feature of many of the models investigated here that they significantly under-predict the number of QSOs in the *ROSAT* sample. For example, the integral log $N$ – log $S$, $N(> S)$, relation predicted by model H (dashed line in Fig. 4) falls significantly below the $N(> S)$ relation for the *ROSAT* sample (filled circles), while providing a good fit to the $N(> S)$ relation for the EMSS (triangles). Based on model H, we would predict only 75 QSOs in the *ROSAT* sample, compared to the 107 observed. From Poisson statistics, this is a $3\sigma$ discrepancy and is consistent with the low KS probability obtained for model H. However, low count predictions for the *ROSAT* sample are not a universal feature of the models considered above. The polynomial evolution model K (represented by the dotted line in Fig. 4) provides a much better fit to the *ROSAT* counts, predicting 119 QSOs in the *ROSAT* sample, consistent with the number observed to within $\sim 1\sigma$.

From Fig. 1 it is apparent that much of the discrepancy between the samples for power-law evolution models occurs in the number of QSOs predicted/observed at low redshift. It is therefore possible that the differences in the classification of low-redshift ($z < 0.5$) objects with narrow emission lines between the two surveys (see Section 2.2) could explain some of this discrepancy. To estimate the size of this effect, we performed the maximum likelihood analysis for the power-law evolution model with a redshift cut-off for the combined EMSS and *ROSAT* samples, both excluding the ambiguous EMSS sources (model U) and including the 12 emission-line galaxies identified in the *ROSAT* sample (model V) by Georgantopoulos et al. (1994). As can be seen from the reported fits in Table 3, the exclusion of the AGN classified as 'ambiguous' in the EMSS makes little difference to the overall acceptability of the fit (cf. model H). In contrast, the inclusion of the emission-line galaxies in the *ROSAT* sample does increase the KS probability of the fit from 0.0048 to 0.012. Despite the marginal acceptance of the model under these circumstances, we consider it unlikely that the emission-line galaxies identified in the *ROSAT* sample are related to AGN. As discussed in Section 2.2, all members of this class have spectra characteristic of late-type galaxies (see Georgantopoulos et al. 1994), with strong stellar continua and narrow ($< 500$ km s$^{-1}$) [OII] emission lines whose equivalent widths ($\sim 20$ Å) are consistent with 'normal' emission-line galaxies identified in field galaxy surveys at similar optical magnitude limits (see e.g. Broadhurst et al. 1988). Clearly, better optical spectra will be required to determine the true nature of this population.

The relative success of the polynomial luminosity evolution model over the power-law evolution model (with a redshift cut-off) suggests that the latter model is no longer an adequate representation of QSO evolution in the X-ray regime. Nevertheless, the power-law evolution model does provide a good fit to both the EMSS and *ROSAT* samples when they are analysed separately (see above). It is therefore possible that the origin for the discrepancy lies not in the inadequacy of the model but in a systematic effect between the EMSS and *ROSAT* samples (possibly introduced by the different energy ranges for the two samples) which has not been accounted for in the analysis above. The most straightforward origin for any such effect lies in an error in the flux scales for one or both of the samples used in this analysis. However, a power-law luminosity evolution model can only be made compatible with both the EMSS and *ROSAT* samples if no correction is applied to the *ROSAT* 0.5–2 keV fluxes in order to convert to the EMSS 0.3–3.5 keV band, although such models (W and X in Table 3) are only acceptable at the $\gtrsim 1$ per cent level. Since even a radical alteration in the assumed X-ray spectral index for QSOs cannot significantly alter the bandpass correction (see Boyle et al. 1993), such a result would require an $\sim 80$ per cent relative shift between the *Einstein* and *ROSAT* flux scales (with the current *ROSAT* scale being too bright). While residual errors at less than the 20 per cent level may still occur in the flux scales, it is unlikely that an error of this magnitude is still present. Different spectral indices in the rest-frame spectra of the EMSS and *ROSAT* QSOs (at $\sim 1.5$ keV and $\sim 2.5$ keV respectively) could also account for some of the difference in the derived evolution rates. Although this explanation was explored in Paper I, it appears unlikely that the rest-frame X-ray spectra of QSOs harden sufficiently ($\Delta \alpha_X \gtrsim -0.5$) at $> 2$ keV (Stewart et al. in preparation) for this to be a viable explanation. Alternatively, because the *ROSAT* QSOs are mostly observed at higher *rest-frame* energies than the EMSS QSOs, a larger dispersion in $\alpha_X$ at harder energies would decrease the power-law evolution rate derived from the *ROSAT* sample, forcing it into better agreement with that obtained for the EMSS. The increase in the $\sigma(\alpha_X)$ would, however, need to be dramatic. Based on the results presented in Table 3, the required decrease in the power-law evolution parameter to achieve consistency between the EMSS and *ROSAT* samples ($\Delta k \sim 0.4$) could only be obtained by effectively increasing the dispersion in the spectral index from $\sigma(\alpha_X) = 0$ at 1.5 keV to $\sigma(\alpha_X) = 0.5$ at 2.5 keV.

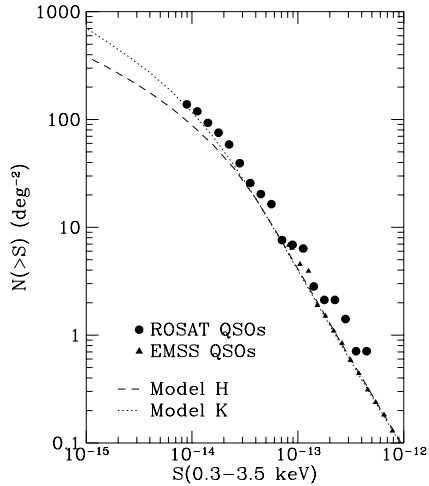

**Figure 4.** The log $N$– log $S$, $N(>S)$, relations for the *ROSAT* sample (filled circles) and EMSS (triangles). The $N(>S)$ predictions based on models H and K (see text) are denoted by the dashed and dotted lines respectively.

It has also recently been claimed by Franceschini et al. (1994) that significant incompleteness exists in the EMSS at flux levels $S(0.3 - 3.5\,\text{keV}) < 2.8 \times 10^{-13}\,\text{erg s}^{-1}\text{cm}^{-2}$. By excluding QSOs in the EMSS with $S(0.3 - 3.5\,\text{keV}) < 2.8 \times 10^{-13}\,\text{erg s}^{-1}\text{cm}^{-2}$, Franceschini et al. (1994) were able to show that the remaining QSOs were consistent with a much higher rate of evolution $L_X \propto (1+z)^{3.5}$, for a spectral index $\alpha_X = 1.2$. This is equivalent to $L_X \propto (1+z)^{3.3}$ for $\alpha_X = 1.0$ adopted here. We have carried out a maximum likelihood analysis on the combined *ROSAT* and EMSS data-sets, with the EMSS restricted to flux levels $S(0.3 - 3.5\,\text{keV}) > 2.8 \times 10^{-13}\,\text{erg s}^{-1}\text{cm}^{-2}$. At this flux limit the EMSS contains 289 QSOs. The best-fit parameters for a power-law evolution model with a redshift cut-off applied to this data-set are listed in Table 3 (models Y and Z) for both $q_0 = 0$ and $q_0 = 0.5$. We find that the KS probabilities of the fits are significantly improved over the equivalent models (G and H) fitted to the entire EMSS, with the KS probability for the $q_0 = 0.5$ model over $\sim 5$ per cent. The evolution rate derived for the $q_0 = 0.5$ model, $k = 3.13\pm 0.1$, is consistent with the value derived by Franceschini et al. (1994), although precise comparison is difficult as the model fitted here includes a redshift cut-off at $z_{\text{max}}$.

Finally, Hasinger et al. (1993) find some evidence for hardening of source spectra in their Lockman Hole survey at fluxes below $1.0 \times 10^{-14}\,\text{erg s}^{-1}\text{cm}^{-2}$. Any hardening of the QSO spectra (and consequent decrease in the derived value of $k$) at such faint flux levels in the *ROSAT* sample could therefore help to explain some of the discrepancy between the EMSS and *ROSAT* evolution rates. However, we find no evidence for any significant increase in the QSO hardness ratios over the 0.5–2 keV band at similar flux levels in our survey (Almaini, private communication). It is therefore probable that sources other than QSOs are responsible for the hardening seen in the Hasinger et al. (1993) sample, which is currently largely unidentified.

### 4.2 The 1–2 keV X-ray background

We can also compute the expected contribution of QSOs to the X-ray background (XRB) based on the observed extragalactic 1–2 keV background intensity, $I_{\text{XRB}}(1\text{–}2\,\text{keV}) = 1.25 \times 10^{-8}\,\text{erg cm}^{-2}\text{s}^{-1}\text{sr}^{-1}$, obtained by Hasinger et al. (1993). Despite the relatively poor fit of the XLF to the data, we may still obtain some useful limits based on the best of these models, since many of the parameters derived in this paper (particularly the evolution at high redshift) are more tightly constrained than in Paper I. We compute the QSO contribution ($I_Q$) to the 1–2 keV XRB by integrating over the $z = 0$ XLF as follows:

$$I_Q = \frac{0.28\,c}{4\pi H_0} \int_{L_X(0)_{\text{min}}}^{L_X(0)_{\text{max}}} \int_{z_{\text{min}}}^{z_{\text{max}}} \frac{L_X \Phi_X(L_X,0)(1+z)^k}{\sqrt{(1+2q_0 z)}(1+z)^{2+\alpha_X}}\,dz\,dL$$

where the integration proceeds over the un-evolved luminosity range $10^{37} < L_X(0) < 10^{47}\,\text{erg s}^{-1}$ and over the redshift range $0 < z < 4$. The factor 0.28 converts the background derived from the 0.3–3.5 keV band XLF to the 1–2 keV band, assuming a spectral index of $\alpha_X = 1$.

Table 4 lists the values of $I_Q$ obtained for five of the best-fit evolutionary models in this paper (models G, H, K, S and T). Beside each estimate of $I_Q$ in this table, we also give the corresponding fractional contribution of QSOs to the 1–2 keV XRB. The QSO contribution ranges from $0.42 \times 10^{-8} - 0.66 \times 10^{-8}\,\text{keV cm}^{-2}\text{s}^{-1}\text{sr}^{-1}$ or 34–53 per cent of the 1–2 keV XRB. The best-fit polynomial evolution model to the combined *ROSAT* and EMSS data-set (model K, $q_0 = 0$) predicts a contribution of 50 per cent to the 1–2 keV XRB. The range is lower than that obtained in Paper I, partly because it has become clearer that a redshift cut-off is required in the evolution of QSOs at high redshifts, but principally because the faint-end slopes of the XLF derived in this paper are flatter than those obtained in Paper I (e.g. $\gamma_1 = 1.53$ for $q_0 = 0$ in models G and S, cf. $\gamma_1 = 1.71$ for the corresponding models in Paper I). Nevertheless, the dominant uncertainty in the QSO contribution to the XRB is still the faint-end slope of the XLF. For example, a steepening of the XLF faint-end slope in model T by 0.15 (the $1\sigma$ error) to $\gamma_1 = 1.51$, whilst retaining the same values for the other parameters, would result in a prediction of 62 per cent for the QSO contribution to the XRB. Conversely, a flattening of the slope by the same amount to $\gamma_1 = 1.21$ would give only 33 per cent. In contrast, an increase in the maximum redshift for the integration of the XLF from $z = 3$ to $z = 5$ in model T only results in an increase for the QSO contribution to the XRB from 42 per cent to 46 per cent.

A recent observation of the XRB with the *ASCA* satellite by Gendreau et al. (1994) yields a 1–2 keV background, $I_{\text{XRB}}(1\text{–}2\,\text{keV}) \sim 1.21 \times 10^{-8}\,\text{erg cm}^{-2}\text{s}^{-1}\text{sr}^{-1}$ (based on a normalization of $I_{XRB}(1\,\mathrm{keV}) = 1.39 \times 10^{-8}$ erg cm$^{-2}$s$^{-1}$sr$^{-1}$, and a spectral index $\alpha_X = -0.35$, see table 1 in Gendreau et al. 1994), only 4 per cent less than that derived by Hasinger et al. (1993). This is despite the significantly flatter slope ($\nu^{-0.35}$) measured for the XRB in this energy range by Gendreau et al. (1994). Based on these observations, the QSO contribution to the 1–2 keV background would be correspondingly increased by 4 per cent for all models listed in Table 4.

**Table 4.** The QSO contribution to the 1–2 keV X-ray background.

| Model | $I_Q$ (keV cm$^{-2}$s$^{-1}$sr$^{-1}$) | $I_Q/I_{XRB}$ |
|---|---|---|
| G | $0.64 \times 10^{-8}$ | 0.52 |
| H | $0.42 \times 10^{-8}$ | 0.34 |
| K | $0.63 \times 10^{-8}$ | 0.50 |
| S | $0.66 \times 10^{-8}$ | 0.53 |
| T | $0.56 \times 10^{-8}$ | 0.44 |

## 5 CONCLUSIONS

We have investigated the evolution of the XLF based on a new sample of 107 *ROSAT* X-ray QSOs identified at faint X-ray flux levels, $S(0.5-2\,\mathrm{keV}) > 4 \times 10^{-15}$ erg s$^{-1}$cm$^{-2}$. For $q_0 = 0.5$, the evolution of the sample is consistent at the 95 per cent confidence level with power-law luminosity evolution, $L_X^*(z) \propto (1+z)^{3.25}$ at $z < 1.6$ and a constant comoving space density at higher redshifts. This is significantly higher than the rate of evolution derived previously for the EMSS. Indeed, with the exception of a polynomial evolution model ($L_X^*(z) \propto 10^{(1.14z-0.23z^2)}$, $q_0 = 0$) and a power-law evolution model ($L_X^*(z) \propto (1+z)^{2.5}$, $z_{max} = 1.25$, $q_0 = 0.5$), which also requires a large dispersion in X-ray spectral index, $\sigma(\alpha_X) = 0.5$, most pure luminosity evolution models do not provide a good fit to the combined *ROSAT* and EMSS QSO samples. More complex evolutionary forms (luminosity and density evolution, a redshift-dependent faint-end slope of the XLF) also fail to provide an adequate overall fit to the combined data-set. It is possible that systematic effects introduced by the different energy bands sampled by the EMSS and *ROSAT* surveys (e.g. an increase in the X-ray spectral index dispersion with energy) and/or possible incompleteness in the EMSS could account (at least in part) for the poor fit of many models to the combined EMSS and *ROSAT* data-set. Based on extrapolation of the best-fit models for the XLF and its evolution, the most likely values for the QSO contribution to the 1–2 keV XRB lie in the range 34–53 per cent.

## ACKNOWLEDGMENTS

BJB acknowledges the receipt of a Royal Society University Research Fellowship.

## REFERENCES

Avni Y., Bahcall J.N., 1980, ApJ, 235, 694
Boyle B.J., Shanks T., Peterson B.A., 1988, MNRAS, 238, 935
Boyle B.J., Fong R., Shanks T., Peterson B.A., 1990, MNRAS, 243, 1
Boyle B.J., Griffiths,R.E., Shanks T., Stewart,G.C. Georgantopoulos I., 1993, MNRAS, 260, 49 (Paper I)
Branduardi-Raymont et al., 1993, in Chincarini,D., et al., eds, ASP Conf., Ser., 51, Observational Cosmology, Astron., Soc., Pac., San Francisco, p.466
Broadhurst T. J., Ellis R. S., Shanks T., 1988, MNRAS 235, 827
Della Ceca R., Maccacaro T., Gioia I.M., Wolter A. Stocke J.T., 1992, ApJ, 389, 491
Dunlop J. A., Peacock J. A., 1990, MNRAS, 247, 19
Franceschini A., La Franca F., Cristiani S. Martin-Mirones,J.M., 1994, MNRAS, 269, 683
Francis P. J., 1993, ApJ, 407, 519
Gendreau K. G., et al., 1994, preprint
Georgantopoulos I. G., Stewart G. C., Shanks T., Griffiths R. E., Boyle B. J., 1994, MNRAS, submitted
Giallongo E., Vagnetti F., 1992, ApJ, 396, 411
Hasinger G., Burg R., Giacconi R., Hartner G., Schmidt M. Trümper J., Zamorani G., 1993, A&A, 275, 1
Maccacaro T., Della Ceca R., Gioia I.M., Morris S.L. Stocke J.T., Wolter A., 1991, ApJ, 374, 117 (M91)
Marshall H.L., Avni Y., Braccesi A., Huchra J., Tananbaum H. Zamorani G., Zitelli V., 1984, ApJ, 283, 50
Peacock J.A., 1985, MNRAS, 217, 601
Press W. H., Teukolsky S. A., Vetterling W. T., Flannery B. P., 1992, Numerical Recipes in Fortran, CUP, Cambridge, p.640
Shanks T., Georgantopoulos I., Stewart G. C., Pounds K. A. Boyle B. J., Griffiths R. E., 1991, Nat, 353, 315
Stocke J. T., Morris S. L., Gioia I. M., Maccacaro T. Schild R., Wolter A., Fleming T. A., Henry J. P., 1991, ApJS, 76, 813
Wilkes B. J., Elvis M., 1987, ApJ, 323, 243